\documentclass[12pt]{article}
\usepackage{amsfonts}
\usepackage{amsmath}
\usepackage{amssymb}
\usepackage{myart}

\normalbaselines
\font\tenbm=cmmib10
\font\sevenbm=cmmib7

\newfam\bmfam
\textfont\bmfam=\tenbm \scriptfont\bmfam=\sevenbm
\input tcilatex.tex
\begin{document}

\author{Yuri A. Rylov}
\title{Multivariance as a crucial property of microcosm}
\date{Institute for Problems in Mechanics, Russian Academy of Sciences \\
101-1 ,Vernadskii Ave., Moscow, 117526, Russia \\
email: rylov@ipmnet.ru\\
Web site: {$http://rsfq1.physics.sunysb.edu/\symbol{126}rylov/yrylov.htm$}\\
or mirror Web site: {$http://gasdin-ipm.ipmnet.ru/\symbol{126}%
rylov/yrylov.htm$}}
\maketitle

\begin{abstract}
The conventional method of a generalized geometry construction, based on
deduction of all propositions of the geometry from axioms, appears to be
imperfect in the sense, that multivariant geometries cannot be constructed
by means of this method. Multivariant geometry is such a geometry, where at
the point $P$ there are many vectors $\mathbf{PP}^{\prime }$, $\mathbf{PP}%
^{\prime \prime }$,... which are equivalent to the vector $\mathbf{QQ}%
^{\prime }$ at the point $Q$, but they are not equivalent between
themselves. In the conventional (Euclidean) method the equivalence relation
is transitive, whereas in a multivariant geometry the equivalence relation
is intransitive, in general. It is a reason, why the multivariant geometries
cannot be deduced from a system of axioms. The space-time geometry in
microcosm is multivariant. Multivariant geometry is a grainy geometry, i.e.
the geometry, which is partly continuous and partly discrete. Multivariance
is a mathematical method of the graininess description. The graininess (and
multivariance) of the space-time geometry generates a multivariant (quantum)
motion of particles in microcosm. Besides, the grainy space-time generates
some discrimination mechanism, responsible for discrete parameters (mass,
charge, spin) of elementary particles. Dynamics of particles appears to be
determined completely by properties of the grainy space-time geometry. The
quantum principles appear to be needless.
\end{abstract}

\section{Introduction}

At first about the term "crucial" with respect to a property of a physical
theory. In fifteenth an sixteenth centuries, when the transition from the
Aristotelian mechanics to the Newtonian mechanics took place, the crucial
concept was "inertia". This concept was absent in the Aristotelian
mechanics, but this concept was the new concept, appeared in the Newtonian
mechanics. Formally the increase of the order of dynamic equations of the
physical body motion was connected with the concept of inertia. A chariot
moving by a horse is a symbol of the Aristotelian mechanics (there is no
inertia). A pendulum, whose vibration can be explained only by means of the
concept of inertia, is a symbol of the Newtonian mechanics. Introduction of
the crucial concept into mechanics lasted longer, than a century. This
introduction was accompanied by difficulties and conflicts between the
investigators. For instance, conflict between Ptolemaic successors and
successors of Copernicus was conditioned by a use of the concept of inertia.
According to Ptolemaic conception the planetary system is a great mechanism,
which was put in motion by God, whereas according to Copernicus conception
the planets move themselves by inertia. Finally, the concept of inertia was
so important, that Sir Isaac Newton devoted the first law of mechanics to
formulation of this concept, although in reality the first law of mechanics
is simply a special case of the second law of mechanics. Appearance of the
concept of inertia is conditioned by the transition from earthen mechanics,
where the friction force was a dominating reason of dynamics, to the
celestial mechanics, where the friction force may be neglected.

At transition from the macroscopic mechanics to mechanics of microcosm a new
crucial concept appears. This new concept is called multivariance. When one
investigated a passage of electrons through a narrow slit, one discovered,
that the electron motion ceases to deterministic (electron diffraction). The
electron motion becomes to be multivariant (nondeterministic). Principles of
the classical mechanics do not admit a multivariant motion of a free
particle. However, the experiment shows, that the motion of small
(elementary) particles may be multivariant. Motion is determined by two
factors: (1) space-time geometry and (2) laws of dynamics. Thus, there are
two possibilities: either the space-time geometry is multivariant, or
dynamics in microcosm is multivariant (it may be also, that both geometry
and dynamics are multivariant). In the thirtieth of the twentieth century,
when the electron diffraction was discovered, the multivariant geometry was
not known. Nobody could imagine, that the space-time geometry may be
multivariant. (Note, that nondeterministic geometries were known. But in
reality, there were stochastic structures, given on a geometry, whereas the
geometry in itself was deterministic and single-variant). As a result the
multivariance has been ascribed to dynamics. This multivariant dynamics is
known as the quantum mechanics. Note that appearance of the multivariant
geometry in dynamics is not identical to quantum dynamics. The quantum
dynamics is a special case of multivariant dynamics. The multivariant
dynamics contains the quantum dynamics and something else, which cannot be
reduced to the quantum dynamics. This "something else" is of interest.

It appeared that the geometry may be multivariant \cite{R2001}.
Multivariance of geometry means as follows. Let $\mathbf{P}_{0}\mathbf{P}%
_{1} $ be a vector at the point $P_{0}$. Let at the point $Q_{0}$ there be
many vectors $\mathbf{Q}_{0}\mathbf{Q}_{1}$,$\mathbf{Q}_{0}\mathbf{Q}_{2}$%
,... , which are equivalent (equal) to the vector $\mathbf{P}_{0}\mathbf{P}%
_{1}$ at the point $P_{0}$, but vectors $\mathbf{Q}_{0}\mathbf{Q}_{1}$,$%
\mathbf{Q}_{0}\mathbf{Q}_{2}$,... are not equivalent between themselves. If
such a situation takes place in geometry, then such a geometry is
multivariant.

If at any point $Q_{0}$ there is one and only one vector $\mathbf{Q}_{0}%
\mathbf{Q}_{1}$, which is equivalent to the vector $\mathbf{P}_{0}\mathbf{P}%
_{1}$ at the point $P_{0}$, such a geometry is called the single-variant
geometry.

In general, multivariance and single-variance of the geometry are considered
with respect to some pair of points $P_{0},Q_{0}$. It is possible such a
situation, when the geometry is multivariant with respect to some pairs of
points, and it is single-variant with respect to another pairs of points. If
the geometry is multivariant with respect to at least one pair of points,
such a geometry will be qualified as multivariant. In the multivariant
space-time geometry the particle dynamics appears to be multivariant, even
if this dynamics acts in accordance with conventional principles of the
classical dynamics.

Note, that the equivalence relation is supposed to be transitive in all
mathematical models. i.e. in all logical constructions, which can be deduced
from a system of axioms by means of the rules of the formal logic. The
equivalence relation is transitive by definition, if for any objects (for
instance, vectors $\mathbf{P}_{0}\mathbf{P}_{1}$,$\mathbf{Q}_{0}\mathbf{Q}%
_{1}$,$\mathbf{Q}_{0}\mathbf{Q}_{2}$) it follows from the relations $\mathbf{%
P}_{0}\mathbf{P}_{1}$eqv$\mathbf{Q}_{0}\mathbf{Q}_{1}$ and $\mathbf{P}_{0}%
\mathbf{P}_{1}$eqv$\mathbf{Q}_{0}\mathbf{Q}_{2}$, that $\mathbf{Q}_{0}%
\mathbf{Q}_{1}$eqv$\mathbf{Q}_{0}\mathbf{Q}_{2}$. Here designation "eqv"
means relation of equivalence. Comparison of definition of multivariance ($%
\mathbf{P}_{0}\mathbf{P}_{1}\mathrm{eqv}\mathbf{Q}_{0}\mathbf{Q}_{1}$ $%
\wedge $ $\mathbf{P}_{0}\mathbf{P}_{1}\mathrm{eqv}\mathbf{Q}_{0}\mathbf{Q}%
_{2}$, but $\mathbf{Q}_{0}\mathbf{Q}_{1}\overline{\mathrm{eqv}}\mathbf{Q}_{0}%
\mathbf{Q}_{2}$) with the definition of transitivity shows that the
equivalence relation in the multivariant geometry cannot be always
transitive.

However, does the multivariant geometry (T-geometry) exist? If yes, then how
can one construct a multivariant geometry?

Let us consider the proper Euclidean geometry and define equivalence of two
vectors $\mathbf{P}_{0}\mathbf{P}_{1}$ and $\mathbf{Q}_{0}\mathbf{Q}_{1}$ as
follows. Vectors $\mathbf{P}_{0}\mathbf{P}_{1}$ and $\mathbf{Q}_{0}\mathbf{Q}%
_{1}$ are equivalent ($\mathbf{P}_{0}\mathbf{P}_{1}$eqv \newline
$\mathbf{Q}_{0}\mathbf{Q}_{1}$), if vectors $\mathbf{P}_{0}\mathbf{P}_{1}$
and $\mathbf{Q}_{0}\mathbf{Q}_{1}$ are in parallel $\left( \mathbf{P}_{0}%
\mathbf{P}_{1}\uparrow \uparrow \mathbf{Q}_{0}\mathbf{Q}_{1}\right) $ and
their lengths $\left\vert \mathbf{P}_{0}\mathbf{P}_{1}\right\vert $ and $%
\left\vert \mathbf{Q}_{0}\mathbf{Q}_{1}\right\vert $ are equal.
Mathematically these two conditions are written in the form%
\begin{equation}
\left( \mathbf{P}_{0}\mathbf{P}_{1}\uparrow \uparrow \mathbf{Q}_{0}\mathbf{Q}%
_{1}\right) :\qquad \left( \mathbf{P}_{0}\mathbf{P}_{1}.\mathbf{Q}_{0}%
\mathbf{Q}_{1}\right) =\left\vert \mathbf{P}_{0}\mathbf{P}_{1}\right\vert
\cdot \left\vert \mathbf{Q}_{0}\mathbf{Q}_{1}\right\vert  \label{a1.1}
\end{equation}

\begin{equation}
\left\vert \mathbf{P}_{0}\mathbf{P}_{1}\right\vert =\left\vert \mathbf{Q}_{0}%
\mathbf{Q}_{1}\right\vert ,\qquad \left\vert \mathbf{P}_{0}\mathbf{P}%
_{1}\right\vert =\sqrt{2\sigma \left( P_{0},P_{1}\right) }  \label{a1.2}
\end{equation}%
where $\left( \mathbf{P}_{0}\mathbf{P}_{1}.\mathbf{Q}_{0}\mathbf{Q}%
_{1}\right) $ is the scalar product of two vectors, defined by the relation%
\begin{equation}
\left( \mathbf{P}_{0}\mathbf{P}_{1}.\mathbf{Q}_{0}\mathbf{Q}_{1}\right)
=\sigma \left( P_{0},Q_{1}\right) +\sigma \left( P_{1},Q_{0}\right) -\sigma
\left( P_{0},Q_{0}\right) -\sigma \left( P_{1},Q_{1}\right)  \label{a1.3}
\end{equation}%
Here $\sigma $ is the world function of the proper Euclidean space, which is
defined via the Euclidean distance $\rho \left( P,Q\right) $ between the
points $P,Q$ by means of the relation%
\begin{equation}
\sigma \left( P,Q\right) =\frac{1}{2}\rho ^{2}\left( P,Q\right)  \label{a1.4}
\end{equation}%
The length $\left\vert \mathbf{PQ}\right\vert $ of vector $\mathbf{PQ}$ is
defined by the relation%
\begin{equation}
\left\vert \mathbf{PQ}\right\vert =\rho \left( P,Q\right) =\sqrt{2\sigma
\left( P,Q\right) }  \label{a1.5}
\end{equation}

Using relations (\ref{a1.1}) - (\ref{a1.5}), one can write the equivalence
condition in the form $\mathbf{P}_{0}\mathbf{P}_{1}$eqv$\mathbf{Q}_{0}%
\mathbf{Q}_{1}:$%
\begin{eqnarray}
\sigma \left( P_{0},Q_{1}\right) +\sigma \left( P_{1},Q_{0}\right) -\sigma
\left( P_{0},Q_{0}\right) -\sigma \left( P_{1},Q_{1}\right) &=&\sigma \left(
P_{0},P_{1}\right)  \label{a1.6} \\
\wedge \sigma \left( P_{0},P_{1}\right) &=&\sigma \left( Q_{0},Q_{1}\right)
\label{a1.7}
\end{eqnarray}

The definition of equivalence (\ref{a1.6}), (\ref{a1.7}) is a satisfactory
geometrical definition, because it does not contain a reference to the
dimension of the space and to the coordinate system. It contains only points 
$P_{0},P_{1},Q_{0},Q_{1}$, determining vectors $\mathbf{P}_{0}\mathbf{P}_{1}$
and $\mathbf{Q}_{0}\mathbf{Q}_{1}$ and distances (world functions) between
these points. The definition of equivalence (\ref{a1.6}), (\ref{a1.7})
coincides with the conventional definition of two vectors equivalence in the
proper Euclidean geometry. If one fixes points $P_{0},P_{1},Q_{0}$ in the
relations (\ref{a1.6}), (\ref{a1.7}) and solve them with respect to the
point $Q_{1}$, one finds that these equations always have one and only one
solution. This statement follows from the properties of the world function
of the proper Euclidean geometry. It means that the proper Euclidean
geometry is single-variant with respect any pairs of its points. It means
also, that the equivalence relation is transitive in the proper Euclidean
geometry.

Any geometry is a set (in general, continual one) of propositions. The
proper Euclidean geometry may be axiomatized, i.e. all propositions of the
proper Euclidean geometry may be deduced from a finite set of propositions
(axioms) by means of the rules of formal logic. The system of axioms is
consistent \cite{H30}. This fact is in accordance with the transitivity of
the equivalence relation in the proper Euclidean geometry.

On the other hand, all propositions of the proper Euclidean geometry may be
expressed in terms of the world function \cite{R2001}. Let us represent all
propositions of the proper Euclidean geometry in terms of the Euclidean
world function $\sigma _{\mathrm{E}}$ and replace the Euclidean world
function $\sigma _{\mathrm{E}}$ by some another world function $\sigma $,
satisfying the constraints 
\begin{equation}
\sigma :\quad \Omega \times \Omega \rightarrow \mathbb{R},\qquad \sigma
\left( P,Q\right) =\sigma \left( Q,P\right) ,\qquad \sigma \left( P,P\right)
=0,\qquad \forall P,Q\in \Omega  \label{a1.8}
\end{equation}%
where $\Omega $ is the set of all points, where the geometry is given. We
obtain the set of all propositions of the geometry $\mathcal{G}$, described
by the world function $\sigma $. Such a replacement is a deformation of the
proper Euclidean geometry. Thus, it is possible to construct the "metric"
geometry, which contains all propositions of the proper Euclidean geometry.
I shall not use the term "metric geometry" for the deformed geometry $%
\mathcal{G}$, because the geometry $\mathcal{G}$ is free of the constraint
(the triangle axiom), which is imposed on the metric geometry

\begin{equation}
\rho \left( P,R\right) +\rho \left( R,Q\right) \geq \rho \left( P,Q\right)
,\qquad \forall P,Q,R\in \Omega  \label{a1.9}
\end{equation}

The triangle axiom (\ref{a1.9}) is imposed, in order to conserve
one-dimensional character of shortest (straight) in the metric geometry.
Indeed, in the proper Euclidean geometry the set 
\begin{equation}
\mathcal{EL}_{P_{1},P_{2},Q,}=\left\{ R|\rho \left( P_{1},R\right) +\rho
\left( R,P_{2}\right) =\rho \left( P_{1},Q\right) +\rho \left(
Q,P_{2}\right) \right\}  \label{a1.9a}
\end{equation}%
is an ellipsoid with focuses at the points $P_{1},P_{2}$ and the point $Q$
on the surface of the ellipsoid. If the point $Q$ tends to the focus $P_{2}$%
, the ellipsoid degenerates in segment%
\begin{equation}
\mathcal{T}_{[P_{1},P_{2},]}=\left\{ R|\rho \left( P_{1},R\right) +\rho
\left( R,P_{2}\right) =\rho \left( P_{1},P_{2}\right) \right\}  \label{a1.9c}
\end{equation}%
of the straight line, passing through the points $P_{1}$ and $P_{2}$. In the
proper Euclidean geometry the ellipsoid degenerates into one-dimensional
segment of straight. However, in the arbitrary metric geometry, given on $n$%
-dimensional manifold, the equation 
\begin{equation}
\mathcal{S}:\quad \Phi \left( R\right) =0,\qquad \Phi \left( R\right) \equiv
\rho \left( P_{1},R\right) +\rho \left( R,P_{2}\right) -\rho \left(
P_{1},P_{2}\right)  \label{a1.9b}
\end{equation}%
determines, in general, $\left( n-1\right) $-dimensional closed surface $%
\mathcal{S}$. The points $R$, satisfying the condition $\Phi \left( R\right)
>0$ are external points, which are placed outside the closed surface $%
\mathcal{S}$. The points $R$, satisfying the condition $\Phi \left( R\right)
<0$ are internal points, which are placed inside $\mathcal{S}$. If the
condition (\ref{a1.9}) is satisfied, it means, that the closed surface $%
\mathcal{S}$ has no internal points. In this case the segment (\ref{a1.9c})
has no internal points, i.e. it is one-dimensional.

In the deformed geometry $\mathcal{G}$ the solution of equations (\ref{a1.6}%
), (\ref{a1.7}) for the point $Q_{1}$ at fixed points $P_{0},P_{1},Q_{0}$
does not always exist. If it exists, it is not always unique. In other
words, the deformed geometry $\mathcal{G}$ is multivariant, in general. In
the same time any proposition of the proper Euclidean geometry exists in the
deformed geometry $\mathcal{G}$, although it may have another sense, than
the sense, which this proposition has in the proper Euclidean geometry.
Nevertheless, this proposition is the same proposition, formulated in
different geometries.

Some conventional propositions of the proper Euclidean geometry contain
references to the dimension and to the coordinate system, i.e. to the method
of the geometry description. In the conventional (vector) presentation of
the Euclidean geometry, its dimension is considered to be a property of the
geometry in itself, although there are geometries, where the dimension
cannot be introduced, because for introduction of the dimension the world
function must satisfy a series of constraints, which are very restrictive.
In reality, the dimension of geometry is the dimension of the coordinate
system (the number of coordinates), which is used for the geometry
description. The manifold and its dimension is the conventional method of
the geometry description \cite{R2007}, and one cannot separate this method
from the geometry in itself, until one has not alternative method of the
geometry construction.

The deformed geometry $\mathcal{G}$ is a multivariant geometry, which cannot
be axiomatized, in general. It means that the geometry $\mathcal{G}$ is an
example of a theory, which cannot be considered as a conventional
mathematical model, constructed by means of the formal logic on the basis of
some axiomatics. Returning to the multivariance, discovered in the motion of
electrons, one may state, that the problem of multivariance of the electron
motion can be solved on account of a multivariant geometry. The multivariant
space-time geometry looks more reasonable, than the multivariant dynamics in
the single-variant space-time geometry. Indeed, to obtain multivariant
dynamics one is forced to replace principles of classical dynamics by
quantum principles, which looks rather artificial. In the same time the
multivariance is a natural property of the physical geometry (i.e. of the
geometry, described completely by the world function). It depends on the
form of the world function $\sigma $ in the equivalence relations (\ref{a1.6}%
), (\ref{a1.7}), whether or not the geometry $\mathcal{G}$ is multivariant.
Changing the world function of the space-time, one can change a character of
the multivariance. One can choose such a world function of the space-time,
that the conventional classical description of a particle motion coincide
with the description of quantum mechanics \cite{R91}.

Let the space-time geometry be described by the world function 
\begin{equation}
\sigma _{\mathrm{d}}=\sigma _{\mathrm{M}}+d\mathrm{sgn}\left( \sigma _{%
\mathrm{M}}\right) ,\qquad d\equiv \lambda _{0}^{2}=\frac{\hbar }{2bc}=\text{%
const}  \label{a1.10}
\end{equation}%
\begin{equation*}
\mathrm{sgn}\left( x\right) =\left\{ 
\begin{array}{ccc}
1 & \text{if} & x>0 \\ 
0 & \text{if} & x=0 \\ 
-1 & \text{if} & x<0%
\end{array}%
\right.
\end{equation*}%
where $\sigma _{\mathrm{M}}$ is the world function of the Minkowski
space-time, $\hbar $ is the quantum constant, $c$ is the speed of the light
and $b$ is some universal constant. Then the world chain, consisting of
points $...P_{0},P_{1},...P_{k},...$satisfying the relations 
\begin{equation}
\mathbf{P}_{k}\mathbf{P}_{k+1}\mathrm{eqv}\mathbf{P}_{k+1}\mathbf{P}%
_{k+2},\qquad k=...0,1,..  \label{a1.11}
\end{equation}%
describes the motion of a free particle. It appears, that the motion is
multivariant (stochastic) in the space-time with the world function (\ref%
{a1.10}). Statistical description of these multivariant chains coincides
with the quantum description in terms of the Schr\"{o}dinger equation \cite%
{R91}.

Besides, the space-time (\ref{a1.10}) appears to be discrete, because in
this space-time there are no vectors $\mathbf{PQ}$ of the length $\left\vert 
\mathbf{PQ}\right\vert ^{4}\in \left( 0,\lambda _{0}^{4}\right) $.
Discreteness of the space-time seems to be very surprising, because the
space-time is given on the manifold of Minkowski. Conventionally the
discrete space is associated with a grid. The discrete space-time on the
continuous manifold seems to be impossible. This example shows, that a
physical geometry and a continuous manifold, where the geometry is given,
are quite different things. Manifold and its dimension are only attributes
of the vector representation of the Euclidean geometry (i.e. of the
description method) \cite{R2007}, whereas the discreteness of a geometry is
an attribute of the geometry in itself.

In any mathematical model the equivalence relation is transitive. This
property of the mathematical model provides definiteness (single-variance)
for all conclusions, made on the basis of such a mathematical model. If the
equivalence relation is intransitive, the conclusions, made on the basis of
such a model cease to be definite. They becomes multivariant. The logical
construction with the intransitive equivalence relation and, hence, with
multivariant conclusions is not considered to be a mathematical model,
because it is useless, and one cannot make a definite prediction on the
basis of such a model. Besides, such a model cannot be axiomatized, because
the axiomatization supposes a single-variance of conclusions.

I shall refer to models with multivariant (indefinite) predictions as
intransitive models, or multivariant models. Multivariant models appear
automatically, as soon as they use a multivariant space-time geometry. As
far as models of physical phenomena may not ignore space-time geometry, and
the space-time geometry may be multivariant, one cannot avoid a use of
multivariant models of physical phenomena.

Fortunately, a multivariant model can be reduced to a single-variant model,
provided one unites the set of many conclusions, which follows from one
statement into one conclusion. In other words, one considers the set of
different objects as a statistical ensemble. One may work with the
statistical ensemble, considering it as a single object. Then the
multivariant model may cease to be multivariant. It turns into a
single-variant (transitive) model provided, that its objects be statistical
ensembles of original objects. Such a procedure is known as the statistical
description, which deals with statistically averaged objects. Prediction of
the model about statistically averaged objects (statistical ensembles),
which now are objects of the model, may appear to be single-variant, if the
statistical description is produced properly. In other words, a statistical
description, produced properly, transforms a multivariant model into a
single-variant mathematical model.

Procedure of the statistical description is well known. It is used in
different branches of theoretical physics. However, sometimes one obtains
the single-variant mathematical model, dealing with statistically averaged
objects, without knowing that the model deals with statistically averaged
objects. For instance, the gas dynamics model deals with gas particles.
Motion of gas particle describes the mean motion of gas molecules. However,
the gas dynamics equations (as dynamic equations of the continuous medium)
were deduced, before it became known, that the gas consists of molecules.
Besides, there is more detailed statistical description of the gas molecule
motion, based on the gas kinetic theory (Boltzman equation).

The quantum mechanics is a statistical description of the multivariant
particle motion, which is conditioned directly by the multivariant
space-time geometry (\ref{a1.10}). However, the quantum mechanics is not
considered conventionally as a statistical description of a multivariant
particle motion. One considers the quantum mechanics as a corollary of
special quantum principles of dynamics, which are introduced axiomatically.
In this form the quantum mechanics describes very successfully physical
phenomena of atomic physics. Formal technique of quantum mechanics is rather
simple and comfortable. Many investigators like the formalism of quantum
mechanics, and they argue against the quantum mechanics as a statistical
description of multivariantly moving particles.

Something like this, one had more, than hundred years ago with the thermal
phenomena. The heat was explained as a special heat liquid (thermogen),
whose properties are described by the laws of thermodynamics. The axiomatic
thermodynamics explained very well all thermal phenomena. Attempts of
interpretation of the heat as a chaotic molecular motion met objections of
many investigators, who did not believe in existence of molecules. The heat
as a chaotic molecular motion had been accepted, when it became clear that
the thermal fluctuations cannot be explained by the axiomatic
thermodynamics. The thermal fluctuation can be explained only by the
supposition that the heat is a chaotic molecular motion. However, the
axiomatic thermodynamics is much simpler, than the statistical theory of the
heat. It is used now in the theory of continuous medium and other
applications.

Situation with the quantum mechanics looks as follows. In general, the
quantum mechanics may be deduced as a result of statistical description of
the multivariant motion of particles, conditioned by the multivariant
space-time geometry of the form (\ref{a1.10}). However, the quantum
mechanics had been formulated in the beginning of the twentieth century as
an axiomatic conception. The multivariant space-time geometry was not known
then. Now most of investigators do not see a necessity of introducing the
concept of multivariant geometry. The fact, that introduction of quantum
principles is a corollary of our imperfect knowledge of geometry, does not
disturb them. They believe that the relativistic quantum theory and the
theory of elementary particles can be constructed on the basis of
unification of the quantum principles and principles of relativity.

Strategy of further investigations of the microcosm depends essentially on
the approach to the multivariant space-time geometry. If we believe, that
the multivariant space-time geometry is impossible, and quantum principles
reflect the nature of the microcosm, we are forced to use the investigation
strategy, which has been used at the construction of the nonrelativistic
quantum mechanics. The quantum mechanics has been constructed by the
cut-and-try method. The same method is used for further investigation of the
microcosm. Besides, the quantum principles supposes, that all physical
objects and all physical fields are to be quantized. In particular, one
should quantized the gravitational and electromagnetic fields.

On the contrary, if one takes, that the quantum effects are a corollary of
the multivariant geometry, one should not quantized the electromagnetic and
gravitational fields, because these fields describe the space-time geometry.
Besides, the dynamic equations of the electromagnetic field and those of the
gravitational field do not contain the quantum constant. This fact manifests
a distinction of dynamic equations of these fields from the Schr\"{o}dinger
and Dirac equations. From the logical viewpoint the approach, based on a use
of the multivariant space-time geometry, is more consistent also. Indeed,
why would one use only single-variant space-time geometries, which form only
a small part of all possible space-time geometries? When it appears, that
the single-variant space-time cannot explain the multivariant motion of free
particles, one is forced to introduce enigmatic quantum principles to
explain quantum effects, which are a manifestation of multivariance. In
general, why is one to ignore the property of multivariance, which is
observed in experiments on the electron diffraction?

Note that according to the definition of equivalence (\ref{a1.6}), (\ref%
{a1.7}), there exists the zero-variance, when the equations (\ref{a1.6}), (%
\ref{a1.7}) have no solutions. If the multivariance may be reduced to
single-variance of statistically averaged objects by means of statistical
description, the zero-variance of the space-time geometry cannot be
described by a single-variant mathematical model. The zero-variance means
discrimination, when some variants of the particle motion are discriminated.
For instance, the space-time geometry (\ref{a1.10}) discriminates the
particles of small masses, because in the multivariant space-time geometry
the masses of particles are geometrized, and the particle mass $m$ is
connected with the lengths $\left\vert \mathbf{P}_{k}\mathbf{P}%
_{k+1}\right\vert $ of the vectors of the world chain by the relation 
\begin{equation}
m=b\left\vert \mathbf{P}_{k}\mathbf{P}_{k+1}\right\vert  \label{a1.12}
\end{equation}%
where $b$ is the universal constant, which enters in the expression (\ref%
{a1.10}) for the elementary length $\lambda _{0}$.

The fact, that masses of elementary particles, their electric charges and
their internal angular moments (spin) are discrete quantities (but not all
possible ones) is a result of some discrimination mechanism connected with
the possible zero-variance of the space-time geometry. The values of
electric charge and those of spin are multiple to quantities $e$ and $\hbar $
respectively. This fact is postulated in the framework of quantum mechanics.
Discreteness of the elementary particles masses is postulated also. However,
the values of masses are taken from experiment, and theorists dream to
deduce the receipt of determination of the mass values, considering this
receipt as a great achievement of the elementary particle theory. However,
the quantum principles do not admit to determine discrete values of the
elementary particles masses. These discrete values of masses (as well as the
values of the electric charge and spin) should be determined by some
discrimination mechanism, which is conditioned by the multivariant
(zero-variant) space-time geometry. Such a possibility must be investigated,
because, being a corollary of a statistical description, the quantum
principles do not admit one to determine such a discrimination mechanism.

Investigation of the space-time geometry admits one to set the question,
what elementary particles may exist at given space-time geometry. To
determine the proper space-time geometry, one may variate the values of the
world function (\ref{a1.10}) in the interval, where $\sigma \in \left(
-\lambda _{0}^{2},\lambda _{0}^{2}\right) $. Variation of the form of the
world function $\sigma $ for the values of argument $\sigma _{\mathrm{M}}$
in the interval, where $\sigma \in \left( -\lambda _{0}^{2},\lambda
_{0}^{2}\right) $ does not influence on the Schr\"{o}dinger equation,
generated by the multivariant geometry (\ref{a1.10}). In the conventional
approach, when only single-variant space-time geometry is considered, the
question on geometrical justification of the elementary particles existence
cannot be put at all. This question is set only on the level of dynamics,
where one has no discrimination mechanism. In the multivariant space-time
geometry one can consider the question of the limited divisibility of
geometrical objects \cite{R2006}. In the single-variant geometry such a
question cannot be put, because in such a geometry the unlimited
divisibility takes place.

\section{Why do the most scientists ignore concept of multivariant
space-time geometry and concept of multivariance?}

This question is not a scientific question. This is a social near-scientific
question. I do not know the answer for this question. But this question is
very important for further development of the microcosm physics, because it
admits one to choose an effective investigation strategy. I try to consider
different versions of the answer. In reality, one separates this global
question into a series of more special questions and tries to answer some of
them.

It is impossible to find a defect in construction of T-geometry in itself.
It is too simple, in order one could find a mistake or a defect in its
construction. There are three points in the method of construction of
T-geometry:

(1) T-geometry is a physical (metric) geometry, which is described
completely by the world function and only by the world function.

(2) Method of construction of geometrical objects and of the T-geometry
propositions is the same for all T-geometries, i.e. the formula of
description in terms of the world function is the same in all T-geometries.

(3) The proper Euclidean geometry is a mathematical (axiomatized) geometry
and a physical geometry simultaneously. There is a theorem of the Euclidean
geometry, which states, that the proper Euclidean geometry may be described
in terms of the world function and only in terms of the world function \cite%
{R2001}.

The point (2) follows from the point (1). Indeed, let the geometrical object
be described by the formula $a_{1}$ in a physical geometry $\mathcal{G}_{1}$%
, and the same object be described by the formula $a_{2}$ in other physical
geometry $\mathcal{G}_{2}$. If formulas $a_{1}$ and $a_{2}$ are different,
it means that the geometries $\mathcal{G}_{1}$ and $\mathcal{G}_{2}$
distinguish not only by their world functions. There is some quantity, which
is different for $\mathcal{G}_{1}$ and $\mathcal{G}_{2}$, and this
circumstance disagrees with the point (1).

It follows from the point (3), that all propositions of the proper Euclidean
geometry can be deduced from Euclidean axioms and presented in terms of the
proper Euclidean world function $\sigma _{\mathrm{E}}$. Replacing $\sigma _{%
\mathrm{E}}$ in all propositions of the proper Euclidean geometry by the
world function $\sigma $ of the physical geometry $\mathcal{G}$, one obtains
all propositions of the geometry $\mathcal{G}$ and, hence, the physical
geometry $\mathcal{G}$ in itself. The point (3) admits one to construct any
physical geometry, basing on our knowledge of the proper Euclidean geometry.

The non-Euclidean method of the physical geometry construction (the
deformation principle ) \cite{R2007b} is simpler, than the conventional
Euclidean method of the geometry construction, because it does not need a
proof of theorems and a test of the axioms consistency. One may say, that
the conventional method takes from Euclid the intermediate product (method
of the geometry construction), whereas the non-Euclidean method takes from
Euclid his final product (the Euclidean geometry in itself). The
intermediate product needs a further work (proof of theorems), whereas the
final product does not need further work, because all necessary theorems are
supposed to be proved at the stage of the proper Euclidean geometry
construction.

Thus, the deformation principle has not difficulties of the Euclidean
method. Besides, it admits one to construct multivariant geometries.
However, the most mathematicians do not accept the deformation principle.
For instance, the author of this paper submitted a report on the
construction of T-geometry to a geometro-topological seminar in the Moscow
Lomonosov University. The secretary of seminar looked through the presented
paper and said something like that: "How strange geometry! There are no
theorems! Only definitions! I think, that such a geometry is not interesting
for participants of our seminar." The secretary was quite right. The main
activity of geometro-topologists is a formulation of theorems and a proof of
them. Such an activity cannot be applied in the geometry, constructed by
means of the deformation principle.

The secretary of another geometro-topological seminar investigated papers,
submitted for reading of my report. The report was devoted to construction
of T-geometry. Submitted documents contained, in particular, the paper \cite%
{R2005}. Investigation of submitted papers lasted almost a year. It was
finished with the decision: "Participants of the seminar are not ready to
hear the report." Such a decision means, that the participants of the
seminar are not able to argue anything against the T-geometry, but
nevertheless, they cannot accept it. Another examples of negative relation
to the T-geometry construction one can find in \cite{R2005}.

I must note, that there are mathematicians, whose relation to the T-geometry
construction was well-minded. They were participants of the seminar on
"geometry as a whole" in the Moscow Lomonosov University. Reports on the
T-geometry construction were read at this seminar several times. However,
participants of this seminar were not geometro-topologists.

The geometro-topologists construct generalized geometries on the basis of a
topological space and corresponding axiomatics, and the negative relation to
T-geometry may be interpreted in the sense, that accepting T-geometry, one
depreciates papers, based on the conventional (topological) approach to the
construction of generalized geometries. However, I should not like to
interpret the negative reaction of topologists in such a way. I should
prefer to understand objective reasons of negative relation to the
T-geometry.

Expression of Euclidean propositions in terms of the world function supposes
another set $\mathcal{A}_{\mathrm{d}}$ of primary axioms, than the set $%
\mathcal{A}_{\mathrm{c}}$ of primary axioms, which are used usually. For
instance, the set $\mathcal{A}_{\mathrm{c}}$ contains the axiom: "The
straight has no width." The system of primary axioms $\mathcal{A}_{\mathrm{d}%
}$ does not contain this statement. The statement "the straight has no
width." is valid (for the proper Euclidean geometry) in the system of axioms 
$\mathcal{A}_{\mathrm{d}}$, but it is a secondary statement in $\mathcal{A}_{%
\mathrm{d}}$. It is a result of the axiomatics $\mathcal{A}_{\mathrm{d}}$
and of definition of the straight. The definition of the straight $\mathcal{T%
}_{P_{0}P_{1}}$, passing through the points $P_{0},P_{1}$ has the from%
\begin{equation}
\mathcal{T}_{P_{0};P_{0}P_{1}}=\mathcal{T}_{P_{0}P_{1}}=\left\{ R|\mathbf{P}%
_{0}\mathbf{R}\parallel \mathbf{P}_{0}\mathbf{P}_{1}\right\}  \label{a2.0}
\end{equation}%
where the relation of collinearity $\mathbf{P}_{0}\mathbf{P}_{1}\parallel 
\mathbf{P}_{0}\mathbf{R}$ is defined by the relation%
\begin{equation}
\mathbf{P}_{0}\mathbf{R}\parallel \mathbf{P}_{0}\mathbf{P}_{1}\mathbf{%
:\qquad }\left( \mathbf{P}_{0}\mathbf{R}.\mathbf{P}_{0}\mathbf{P}_{1}\right)
^{2}=\left\vert \mathbf{P}_{0}\mathbf{P}_{1}\right\vert ^{2}\left\vert 
\mathbf{P}_{0}\mathbf{R}\right\vert ^{2}  \label{a2.0a}
\end{equation}%
Here the scalar product $\left( \mathbf{P}_{0}\mathbf{P}_{1}.\mathbf{P}_{0}%
\mathbf{R}\right) $ is defined by the relation (\ref{a1.3}). In general, one
equation (\ref{a2.0a}) defines a $(n-1)$-dimensional surface on the $n$%
-dimensional manifold (but not a one-dimensional straight). The statement,
that the point set (\ref{a2.0}), (\ref{a2.0a}) is a one-dimensional
straight, which has no width, follows from the properties of the Euclidean
world function. This property may not take place for other world function.

Note that the point set%
\begin{equation}
\mathcal{T}_{P_{0};Q_{0}Q_{1}}=\left\{ R|\mathbf{P}_{0}\mathbf{R}\parallel 
\mathbf{Q}_{0}\mathbf{Q}_{1}\right\}  \label{a2.0b}
\end{equation}%
\begin{equation}
\mathbf{P}_{0}\mathbf{R}\parallel \mathbf{Q}_{0}\mathbf{Q}_{1}\mathbf{%
:\qquad }\left( \mathbf{P}_{0}\mathbf{R}.\mathbf{Q}_{0}\mathbf{Q}_{1}\right)
^{2}=\left\vert \mathbf{P}_{0}\mathbf{P}_{1}\right\vert ^{2}\left\vert 
\mathbf{P}_{0}\mathbf{R}\right\vert ^{2}  \label{a2.0c}
\end{equation}%
is also a $(n-1)$-dimensional surface on the $n$-dimensional manifold. In
the proper Euclidean geometry the set $\mathcal{T}_{P_{0};Q_{0}Q_{1}}$
degenerates into the straight line, passing through the point $P_{0}$ in
parallel with the vector $\mathbf{Q}_{0}\mathbf{Q}_{1}$ at the point $Q_{0}$.

In the $n$-dimensional Riemannian geometry the $\left( n-1\right) $%
-dimensional point set (\ref{a2.0}), (\ref{a2.0a}) also degenerates into
one-dimensional geodesic, passing to the point $P_{0}$ in parallel with the
vector $\mathbf{P}_{0}\mathbf{P}_{1}$. This degeneration is conditioned by
the fact, that the Riemannian space may be considered as a metric space with
the metric, satisfying the triangle axiom. However, the point set (\ref%
{a2.0b}), (\ref{a2.0c}) does not degenerate, in general, into
one-dimensional curve (geodesic). In the $n$-dimensional Riemannian geometry
the point set (\ref{a2.0b}), (\ref{a2.0c}) remains to be a $\left(
n-1\right) $-dimensional surface, as well as in any T-geometry (except for
the proper Euclidean geometry).

This fact means, that the Riemannian geometry is a multivariant physical
geometry, although some sort of straights ($\mathcal{T}_{P_{0};P_{0}P_{1}}=%
\mathcal{T}_{P_{0}P_{1}}$) is single-variant (one-dimensi\-onal). On the
other hand, the Riemannian geometry is constructed usually as a
single-variant geometry, and existence of geometrical objects (\ref{a2.0b}),
(\ref{a2.0c}) is incompatible with the axiom "geodesic has no width".
Geodesic is defined as a curve of minimal (extremal) length. In turn the
curve is defined as a continuous mapping%
\begin{equation*}
\left[ 0,1\right] \rightarrow \Omega
\end{equation*}%
which cannot be formulated in terms of the world function only, because it
contains a reference to a manifold. To remove the disagreement between the
multivariance of definition (\ref{a2.0b}), (\ref{a2.0c}) and axiom "geodesic
has no width", one declared, that there is no fernparallelism in the
Riemannian geometry, i.e. parallelism of remote vectors is not determined.
At such a constraint the geometrical objects (\ref{a2.0b}), (\ref{a2.0c})
are not defined, and hence, they do not exist.

However, the removal of the fernparallelism does not eliminate inconsistency
of the Riemannian geometry, it eliminates only one of corollaries of this
inconsistency. There may be another (unknown) corollaries of this
inconsistency. One can eliminate these corollaries only removing their
reason (axiom, that the geodesic has no width). It means that one should
accept the definition of the straight (geodesic) in the form (\ref{a2.0b}), (%
\ref{a2.0c}), i.e. the idea of multivariance should be accepted.

Inconsistency of a conception manifests itself only, if one solves a problem
by different correct methods and obtains different results. However, rare
scientists investigate a complicate problem by several different methods and
compare the obtained results.

The outstanding topologist G.Perelman had proved the Poincar\`{e} conjecture 
\cite{P2002,P2003a,P2003b}. In 2006 he was awarded with the Fields medal.
However, he declined to accept the award. He is the only person ever to
refuse the award. Besides, he declined to publish his papers in a peer
review journal, that was necessary for accepting a prize of a million
dollars. His behavior looked strange and unexpected for mathematical
society. Alexander Abramov \cite{A2006}, who knew Perelman personally very
well, describes his style of work as follows. Perelman considered several
versions of solution of the problem and chose the best one. Such a rare
style of investigation is the best one for discovery of inconsistencies in
the Riemannian geometry. Apparently, after publication of his papers in
Archives Perelman has discovered, that the conventional (topological)
approach to the Riemannian geometry is inconsistent (maybe, the paper \cite%
{R2005}, appeared in March 2005, gave a motive for his investigation). But
G.Perelman is a topologist and his papers on the Poincar\`{e} conjecture are
based on the Riemannian geometry. If the Riemannian geometry is
inconsistent, his own papers become questionable, even if all his
considerations are valid.

His further behavior is conditioned by his scientific scrupulosity. He could
not withdraw his papers from Archives, where they were published (it is
prohibited by the rules of Archives). But he could decline publication of
his papers in the peer review mathematical journals. He could not accept the
Fields medal, because some time later his papers may be declared to be
questionable. He should publish the fact, that he discovered inconsistency
of the Riemannian geometry. But such a paper would be a dissident paper.
Anybody, who have written a dissident paper, knows very well, how difficult
to publish a dissident paper. Discussing with colleagues a possible
inconsistency of the Riemannian geometry, G.Perelman did not meet mutual
understanding by colleagues. As a result of such discussions he left the
Institute, where he worked. The charge of his colleagues in scientific
dishonesty is also a result of these discussions.

I did not know G. Perelman personally, and my description of his dignified
behavior is only a hypothesis. But it is a very reasonable hypothesis, which
explains freely all facts by the scientific scrupulosity of G. Perelman and
by his capacity of investigation work. My estimation of the Perelman's
activity distinguishes from position of other scientists, because I know
definitely, that the Riemannian geometry is inconsistent, especially in that
its part, which concerns topology, whereas other scientists cannot admit an
inconsistency of the Riemannian geometry. Topology in the Riemannian
geometry, as well as in other physical geometries, is determined completely
by the world function. The topology may not be given independently, because
in this case one risks to obtain inconsistency.

Construction of multivariant geometry is connected with a replacement of the
formal logic by the "Euclidean logic" \cite{R2005a}, when rules of the
formal logic are substituted be the rules of construction of the Euclidean
geometry propositions in terms of the world function. The transition from
the formal logic to the "Euclidean logic" is a transition from one system of
axioms to another equivalent (for Euclidean geometry) system of axioms. Such
a transition is used very uncommon in the practice of mathematical
investigations. Although possibility of such transition is accepted, but in
practice the transformation of the system of axioms, connected with such a
transition, is used insufficiently. In application of any axiomatics there
are logical stereotypes, when a chain of logical conclusions is replaced by
one statement. Such stereotypes depend on the used axiomatics, and they are
changed at a change of axiomatics. At a replacement of the axiomatics the
logical stereotypes are to be analyzed and replaced by new logical
stereotypes. Unfortunately, the practice of work with logical stereotypes is
insufficient. As a result the old logical stereotypes disturb the perception
of new axiomatics.

Let me adduce a simple example. In the conventional approach to geometry,
based on the vector representation, the discrete geometry cannot be given on
a continual set of points (on manifold). It can be given only on a discrete
set of point of the type of a grid. On the other hand, by definition, the
discrete geometry is such a geometry, where there are no close points. In
the approach, based on the principle of deformation, the distance between
points is determined by the world function and only by the world function.
It is of no importance, where the world function is given (on a grid, or on
a continuous manifold). If the world function is given on a manifold in such
a form, that there are no values of the world function $\sigma $ in the
intervals $\left( -a,0\right) $ and $\left( 0,a\right) ,$ $a>0$, then in the
geometry there are no close points, and the geometry is discrete, even it is
given on a continuous manifold. 

The statement (st): "the discrete geometry cannot be given on a manifold" is
a logical stereotype of the approach, based on the vector representation of
the geometry. This stereotype consists of two statements: (1) definition:
the discrete geometry does not contain close points, (2) axiom: the
continuous geometry is given on a manifold. Although the statement (st) does
not follow strictly from statements (1) and (2) does not follow strictly,
because  it is not known, where the discrete geometry is given.
Nevertheless, because of insufficient development of the discrete geometry
one concludes, that the discrete geometry cannot be given on a manifold, as
far as on a manifold the continuous geometry is given.

I could not overcome the stereotype (st) during almost fifteen years, when I
developed T-geometry. I could not overcome this stereotype, although during 
fifteen years I delt with the discrete geometry, which was described by the
world function (\ref{a4.0}),  given on a continuous manifold. I could not
overcome the stereotype, although I developed the world function formalism
without any problems. I could not overcome the stereotype, although its
essence lies on the surface of the phenomenon. This stereotype is not a
unique one. I met another stereotypes at other scientists. I think, that
such stereotypes do not admit one to accept idea of deformation principle.
In turn the difficulties with overcoming of such stereotypes are connected
with the circumstance, that the transition from one axiomatics to another
equivalent axiomatics is used very rare in practice. The training of
mathematicians for such transitions is too small.

\section{Multivariance and dimension}

Returning to the T-geometry, I should like to manifest, that the concept of
dimension may have different meaning at the conventional approach to
geometry and at the approach based on the deformation principle. I shall
show, that the dimension of geometry and the dimension of the manifold,
where the geometry is given, are different things. The dimension of the
manifold $n_{\mathcal{M}}$ and dimension $n_{\mathcal{G}}$ of the geometry
are different concepts, which coincide for the proper Euclidean geometry.
However, in other physical geometries the two quantities do not coincide, in
general.

Let us consider very simple example of the two-dimensional proper Euclidean
geometry $\mathcal{G}_{\mathrm{E}}$, given on the two-dimensional manifold.
The world function has the form%
\begin{equation}
\sigma _{\mathrm{E}}\left( P_{1},P_{2}\right) =\sigma _{\mathrm{E}}\left( 
\mathbf{x,y}\right) =\frac{1}{2}\left( \left( x^{1}-y^{1}\right) ^{2}+\left(
x^{2}-y^{2}\right) ^{2}\right) ,\qquad \sigma _{\mathrm{E}}\geq 0
\label{a2.1}
\end{equation}%
where the points $P_{0},P_{1}$, $P_{2}$ are three points, whose coordinates
in the Cartesian coordinate system are 
\begin{equation}
P_{0}=\left\{ 0,0\right\} \qquad P_{1}=\left\{ x^{1},x^{2}\right\} ,\qquad
P_{2}=\left\{ y^{1},y^{2}\right\}  \label{a2.2}
\end{equation}%
Vectors $\mathbf{P}_{0}\mathbf{P}_{1}$ and $\mathbf{P}_{0}\mathbf{P}_{2}$
have the Cartesian coordinates%
\begin{equation}
\mathbf{P}_{0}\mathbf{P}_{1}=\mathbf{x}=\left\{ x^{1},x^{2}\right\} ,\qquad 
\mathbf{P}_{0}\mathbf{P}_{2}=\mathbf{y}=\left\{ y^{1},y^{2}\right\}
\label{a2.3}
\end{equation}

Besides, one considers a deformed physical geometry $\mathcal{G}_{\mathrm{d}%
} $, described by the world function%
\begin{equation}
\sigma _{\mathrm{d}}\left( P_{1},P_{2}\right) =\sigma _{\mathrm{E}}\left(
P_{1},P_{2}\right) +d\left( \sigma _{\mathrm{E}}\left( P_{1},P_{2}\right)
\right)  \label{a2.4}
\end{equation}%
where 
\begin{equation}
d\left( \sigma _{\mathrm{E}}\right) =\left\{ 
\begin{array}{ccc}
-\lambda _{0}^{2} & \text{if} & \sigma _{\mathrm{E}}>\sigma _{0} \\ 
-\lambda _{0}^{2}\frac{\sigma _{\mathrm{E}}}{\sigma _{0}} & \text{if} & 
0\leq \sigma _{\mathrm{E}}\leq \sigma _{0}%
\end{array}%
\right. ,\qquad \lambda _{0}^{2}\geq \sigma _{0}\geq 0,\qquad \lambda
_{0},\sigma _{0}=\text{const }  \label{a2.5}
\end{equation}%
Here $\lambda _{0}$ is some elementary length, which is characteristic for
the distorted geometry $\mathcal{G}_{\mathrm{d}}$.

By definition the dimension $n$ of a geometry is the maximal number of
linear independent vectors. In the given case dimension of $\mathcal{G}_{%
\mathrm{E}}$ is equal to $2$, as well as the dimension of the manifold,
where the geometry is given. Dimension of the manifold is defined as the
number of coordinates of the coordinate system.

Dimension of the manifold in the physical geometry $\mathcal{G}_{\mathrm{d}}$
is also $2$, as well as in $\mathcal{G}_{\mathrm{E}}$. In the physical
geometry (T-geometry) $m$ vectors $\mathbf{P}_{0}\mathbf{P}_{1},\mathbf{P}%
_{0}\mathbf{P}_{2},...\mathbf{P}_{0}\mathbf{P}_{m}$ are linear independent
if and only if the Gram's determinant 
\begin{equation}
F_{m}\left( \mathcal{P}^{m}\right) \neq 0,\qquad F_{m}\left( \mathcal{P}%
^{m}\right) \equiv \det \left\vert \left\vert \left( \mathbf{P}_{0}\mathbf{P}%
_{i},\mathbf{P}_{0}\mathbf{P}_{k}\right) \right\vert \right\vert ,\qquad
i,k=1,2,...m  \label{a2.6}
\end{equation}%
Here $\mathcal{P}^{m}=\left\{ P_{0},P_{1},...P_{m}\right\} $, and the scalar
product $\left( \mathbf{P}_{0}\mathbf{P}_{i},\mathbf{Q}_{0}\mathbf{Q}%
_{k}\right) $ of two vectors $\mathbf{P}_{0}\mathbf{P}_{1}$, $\mathbf{Q}_{0}%
\mathbf{Q}_{1}$ is defined by the relation (\ref{a1.3}).

The conventional definition of linear independence is different. $s$ vectors 
$\mathbf{P}_{0}\mathbf{P}_{1}$, $\mathbf{P}_{0}\mathbf{P}_{2},...\mathbf{P}%
_{0}\mathbf{P}_{s}$ are linear independent, if the linear combination of
vectors satisfies the relation 
\begin{equation}
\dsum\limits_{k=1}^{k=s}\alpha _{k}\mathbf{P}_{0}\mathbf{P}_{k}=0
\label{a2.8}
\end{equation}%
only at $\alpha _{k}=0,$\ \ \ $k=1,2,...s$. For the proper Euclidean
geometry both definitions (\ref{a2.6}) and (\ref{a2.8}) are equivalent. For
the distorted geometry $\mathcal{G}_{\mathrm{d}}$ they are not equivalent,
in general.

The conventional definition (\ref{a2.8}) supposes existence of the linear
vector space with a scalar product, given on it, and, in particular, it
supposes the procedures of definition : summation of vectors and
multiplication of a vector by a real number. Definition (\ref{a2.6})
contains references only to the world function and points of the space.
Existence of the linear vector space and linear operations over vectors is
not supposed. It is evident, that the definition (\ref{a2.6}) is a more
general definition, than (\ref{a2.8}), which can be applied, only if the
linear vector space can be introduced. It seems rather unexpected, that one
can define linear dependence, without introduction of the linear space,
because the name "linear dependence" implicates conventionally existence of
the linear space. However, the definition (\ref{a2.6}) can be used in the
case, when one cannot introduce the linear space. In this case the
determinant, constructed of scalar products of vectors, describes
interrelations of $m$ vectors, in particular, their mutual orientation.

Let us consider four vectors 
\begin{equation}
\mathbf{P}_{0}\mathbf{P}_{1}=\left\{ a,0\right\} ,\qquad \mathbf{P}_{0}%
\mathbf{P}_{2}=\left\{ 0,b\right\} ,\qquad \mathbf{P}_{0}\mathbf{P}%
_{3}=\left\{ a,b\right\} ,\qquad \mathbf{P}_{0}\mathbf{P}_{2}=\left\{
2a,0\right\}  \label{a2.9}
\end{equation}%
Let us suppose for simplicity, that coordinates $a,b\gg \lambda _{0}$. Then
the scalar products of any vectors (\ref{a2.9}) in the proper Euclidean
geometry $\mathcal{G}_{\mathrm{E}}$ and in the distorted geometry $\mathcal{G%
}_{\mathrm{d}}$ are connected by the relations%
\begin{equation}
\left( \mathbf{P}_{0}\mathbf{P}_{i},\mathbf{P}_{0}\mathbf{P}_{k}\right) _{%
\mathrm{d}}=\left( \mathbf{P}_{0}\mathbf{P}_{i},\mathbf{P}_{0}\mathbf{P}%
_{k}\right) _{\mathrm{E}}-2\lambda _{0}^{2},\qquad \text{if }P_{i}\neq P_{k}
\label{a2.10}
\end{equation}%
\begin{equation}
\left( \mathbf{P}_{0}\mathbf{P}_{i},\mathbf{P}_{0}\mathbf{P}_{i}\right) _{%
\mathrm{d}}=\left( \mathbf{P}_{0}\mathbf{P}_{i},\mathbf{P}_{0}\mathbf{P}%
_{i}\right) _{\mathrm{E}}-\lambda _{0}^{2}  \label{a2.11}
\end{equation}%
These relations are the corollaries of the relations (\ref{a2.4}) and (\ref%
{a1.3})

The Gram's determinant in $\mathcal{G}_{\mathrm{d}}$ for the first three
vectors (\ref{a2.9}) has the form%
\begin{equation}
\left\vert 
\begin{array}{ccc}
a^{2}-2\lambda _{0}^{2} & -\lambda _{0}^{2} & a^{2}-\lambda _{0}^{2} \\ 
-\lambda _{0}^{2} & b^{2}-2\lambda _{0}^{2} & b^{2}-\lambda _{0}^{2} \\ 
a^{2}-\lambda _{0}^{2} & b^{2}-\lambda _{0}^{2} & a^{2}+b^{2}-2\lambda
_{0}^{2}%
\end{array}%
\right\vert =-4\lambda _{0}^{2}\left( b^{2}-\lambda _{0}^{2}\right) \left(
a^{2}-\lambda _{0}^{2}\right)  \label{a2.12}
\end{equation}%
For the four vectors (\ref{a2.9}) the Gram's determinant in $\mathcal{G}_{%
\mathrm{d}}$ has the form 
\begin{eqnarray}
&&\left\vert 
\begin{array}{cccc}
a^{2}-2\lambda _{0}^{2} & -\lambda _{0}^{2} & a^{2}-\lambda _{0}^{2} & 
2a^{2}-\lambda _{0}^{2} \\ 
-\lambda _{0}^{2} & b^{2}-2\lambda _{0}^{2} & b^{2}-\lambda _{0}^{2} & 
b^{2}-\lambda _{0}^{2} \\ 
a^{2}-\lambda _{0}^{2} & b-\lambda _{0}^{2} & a^{2}+b^{2}-2\lambda _{0}^{2}
& 2a^{2}+b^{2}-\lambda _{0}^{2} \\ 
2a^{2}-\lambda _{0}^{2} & b^{2}-\lambda _{0}^{2} & 2a^{2}+b^{2}-\lambda
_{0}^{2} & a^{2}+b^{2}-2\lambda _{0}^{2}%
\end{array}%
\right\vert  \notag \\
&=&-\lambda _{0}^{2}\left( 12a^{4}\lambda
_{0}^{2}-12a^{4}b^{2}+2a^{2}\lambda _{0}^{4}+6b^{2}\lambda _{0}^{4}-5\lambda
_{0}^{6}-3a^{2}b^{2}\lambda _{0}^{2}\right)  \label{a2.14}
\end{eqnarray}

In $\mathcal{G}_{\mathrm{d}}$ the Gram's determinant for two "collinear"
vectors $\mathbf{P}_{0}\mathbf{P}_{1}=\left\{ a,0\right\} ,$ $\mathbf{P}_{0}%
\mathbf{P}_{2}=\left\{ 2a,0\right\} $ has the form.

\begin{equation}
\left\vert 
\begin{array}{cc}
a^{2}-\lambda _{0}^{2} & 2a^{2}-2\lambda _{0}^{2} \\ 
2a^{2}-2\lambda _{0}^{2} & 4a^{2}-\lambda _{0}^{2}%
\end{array}%
\right\vert =3\lambda _{0}^{2}\left( a^{2}-\lambda _{0}^{2}\right)
\label{a2.15a}
\end{equation}%
although in the Euclidean geometry $\mathcal{G}_{\mathrm{E}}$ this
determinant vanishes. In general, in the geometry $\mathcal{G}_{\mathrm{E}}$
all three determinants (\ref{a2.12}), (\ref{a2.14}), (\ref{a2.15a}) vanish,
because $\lambda _{0}^{2}=0$ and dimension of the geometry $\mathcal{G}_{%
\mathrm{E}}$ is equal to $2$.

It follows from (\ref{a2.12}) and (\ref{a2.14}), that in the distorted
geometry $\mathcal{G}_{\mathrm{d}}$ there are, at least, four linear
independent vectors, although the dimension of the manifold remains to be
equal to $2$. One should expect, that in the distorted geometry $\mathcal{G}%
_{\mathrm{d}}$ there is infinite number of linear independent vectors, and
concept of dimension is inadequate for the physical geometry $\mathcal{G}_{%
\mathrm{d}}$.

Thus, in the proper Euclidean geometry the dimension $n_{\mathcal{M}}$ of a
manifold is equal to the dimension $n_{\mathcal{G}}$ of the geometry,
whereas in the distorted geometry $\mathcal{G}_{\mathrm{d}}$ dimension of
the manifold and dimension of the geometry are quite different quantities.
It looks rather unexpected. How is it possible?

The dimension $n_{\mathcal{G}}$ of a geometry is a very complicated concept,
but it concerns to the geometry itself. The dimension $n_{\mathcal{M}}$ of a
manifold is a simple concept, but it relates only to the method of
description (manifold). In the proper Euclidean geometry the values (but not
concepts) of the two dimensions coincide ($n_{\mathcal{G}}=n_{\mathcal{M}}$%
). Conventionally one does not distinguish between the two dimensions. It
leads to a confusion and to an ascription of the description properties to
the geometry in itself.

The dimension $n_{\mathcal{M}}$ of manifold may be defined only for a
continuous set of space points. It is invariant only with respect to
continuous coordinate transformation. In this connection it is interesting a
consideration of the discrete geometry $\mathcal{G}_{\mathrm{dis}}$. Let us
consider the two-dimensional proper Euclidean geometry, given on the point
set $\Omega _{2}$. The point set $\Omega _{2}$ is obtained from the point
set $\Omega $ as follows. Let $K_{2}$ be a Cartesian coordinate system on $%
\Omega $. Let us remove all points of $\Omega $, except of those points, for
which both Cartesian coordinates are integer. The remaining point set $%
\Omega _{2}$ forms a grid. World function $\sigma _{\mathrm{dis}}$ is
defined on the set $\left( \Omega _{2}\times \Omega _{2}\right) \subset
\left( \Omega \times \Omega \right) $. On this set the world function $%
\sigma _{\mathrm{dis}}$ coincides with $\sigma _{\mathrm{E}}$ and, hence, it
satisfies to all conditions of Euclideaness \cite{R2001} except for the
condition IV (the continuity condition). Dimension $n_{\mathcal{G}}$ of
geometry $\mathcal{G}_{\mathrm{dis}}$, determined by means of the definition
(\ref{a2.6}), is equal to $2$. Dimension $n_{\mathcal{M}}$ of the "manifold" 
$\Omega _{2}$ cannot determined definitely, because the number of integer
variables, labeling the points of $\Omega _{2}$ may be $1,2,$...The
dimension $n_{\mathcal{M}}$ is invariant only with respect to continuous
coordinate transformation. In the case, when coordinates are integer, there
are no continuous coordinate transformations. In this case the dimension of
manifold has no sense, because there is no manifold, whereas the geometry
dimension $n_{\mathcal{G}}$ is defined correctly in the case of the discrete
geometry.

\section{Multivariance, discreteness and graininess of \newline
space-time}

Conventionally the discrete geometry is considered on some grid of points.
It seems that a geometry, given on a continuous manifold, cannot be
discrete. It means that conventionally a discreteness of a geometry is
connected with the means of the geometry description, (but not with the
geometry in itself). In reality, the discreteness of the geometry is
determined by the world function. In particular, a discrete geometry may be
given on the continuous manifold. Besides, there may be different degrees of
the physical geometry discreteness.

Let us consider the question on discreteness of the space-time geometry,
described by the world function

\begin{equation}
\sigma _{\mathrm{d}}=\sigma _{\mathrm{M}}+d\cdot \mathrm{sgn}\left( \sigma _{%
\mathrm{M}}\right) ,\qquad d=\lambda _{0}^{2}=\text{const}>0  \label{a4.0}
\end{equation}%
\begin{equation}
\mathrm{sgn}\left( x\right) =\left\{ 
\begin{array}{l}
1,\ \ \text{if}\ \ x>0 \\ 
0,\ \ \ \ \text{if\ \ }x=0 \\ 
-1,\ \ \text{if}\ \ x<0%
\end{array}%
\right. ,  \label{a4.1}
\end{equation}%
where $\sigma _{\mathrm{M}}$ is the world function of the $4$-dimensional
space-time of Minkowski. $\lambda _{0}$ is some elementary length. The
space-time geometry (\ref{a4.0}) is closer to the real space-time geometry
of microcosm, than the space-time of Minkowski, because at this space-time
geometry the quantum effects may be described without a use of the quantum
principles, if the elementary length $\lambda _{0}=\hbar ^{1/2}\left(
2bc\right) ^{-1/2}$. Here $c$ is the speed of the light, $\hbar $ is the
quantum constant, and $b$ is some universal constant, whose exact value is
not determined \cite{R91}.

The space-time geometry (\ref{a4.0}) is a discrete space-time geometry,
because in this space-time geometry there are no vector $\mathbf{P}_{0}%
\mathbf{P}_{1}$, whose length $\left\vert \mathbf{P}_{0}\mathbf{P}%
_{1}\right\vert $ be small enough, i.e. 
\begin{equation}
\left\vert \mathbf{P}_{0}\mathbf{P}_{1}\right\vert ^{4}\notin \left(
0,\lambda _{0}^{4}\right) ,\qquad \forall P_{0},P_{1}\subset \Omega
\label{a4.1a}
\end{equation}%
In other words, the space-time geometry (\ref{a4.0}) has no close points.

Let us consider another space-time geometry $\mathcal{G}_{\mathrm{d}}$ which
is partly discrete. World function $\sigma _{\mathrm{d}}$ of this geometry
has the form 
\begin{equation}
\sigma _{\mathrm{d}}=\sigma _{\mathrm{M}}+d\left( \sigma _{\mathrm{M}}\right)
\label{a4.5}
\end{equation}%
\begin{equation}
d\left( \sigma _{\mathrm{M}}\right) =\lambda _{0}^{2}f\left( \frac{\sigma _{%
\mathrm{M}}}{\sigma _{0}}\right) =\left\{ 
\begin{array}{lll}
\lambda _{0}^{2}\mathrm{sgn}\left( \frac{\sigma _{\mathrm{M}}}{\sigma _{0}}%
\right) & \text{if} & \left\vert \sigma _{\mathrm{M}}\right\vert >\sigma
_{0}>0 \\ 
\lambda _{0}^{2}\frac{\sigma _{\mathrm{M}}}{\sigma _{0}} & \text{if} & 
\left\vert \sigma _{\mathrm{M}}\right\vert \leq \sigma _{0}%
\end{array}%
\right.  \label{a4.6}
\end{equation}%
where $\sigma _{\mathrm{M}}$ is the world function of the geometry of
Minkowski.

If $\sigma _{0}$ is small, the world function is close to the world function
(\ref{a4.0}). If $\sigma _{0}\rightarrow 0$, the world function (\ref{a4.6})
tends to (\ref{a4.0}). Strictly, the space-time geometry (\ref{a4.6}) is not
discrete, however it is close to the discrete space-time geometry (\ref{a4.0}%
).

Let us consider the relative density $\rho \left( \sigma _{\mathrm{d}%
}\right) =d\sigma _{\mathrm{d}}/d\sigma _{\mathrm{E}}$ of points of the
geometry $\mathcal{G}_{\mathrm{d}}$ with respect to the geometry $\mathcal{G}%
_{\mathrm{E}}$. One obtains 
\begin{equation}
\rho \left( \sigma _{\mathrm{d}}\right) =d\sigma _{\mathrm{d}}/d\sigma _{%
\mathrm{E}}=\left\{ 
\begin{array}{ccc}
1 & \text{if} & \left\vert \sigma _{\mathrm{d}}\right\vert >\sigma
_{0}+\lambda _{0}^{2} \\ 
\frac{\sigma _{0}}{\sigma _{0}+\lambda _{0}^{2}} & \text{if} & \left\vert
\sigma _{\mathrm{d}}\right\vert \leq \sigma _{0}+\lambda _{0}^{2}%
\end{array}%
\right.  \label{a4.7}
\end{equation}

If $\sigma _{0}=0$, there is no close points which are separated by interval
with the world function $\sigma _{\mathrm{d}}\in \left( 0,\lambda
_{0}^{2}\right) $ and $\sigma _{\mathrm{d}}\in \left( -\lambda
_{0}^{2},0\right) $. It means, that the space-time geometry is discrete at $%
\sigma _{0}=0$.

If $\sigma _{0}\ll \lambda _{0}^{2}$, the relative density $\rho \left(
\sigma _{\mathrm{d}}\right) \simeq \sigma _{0}/\lambda _{0}^{2}$ of points
inside the interval $\sigma _{\mathrm{d}}\in \left( -\sigma _{0}-\lambda
_{0}^{2},\sigma _{0}+\lambda _{0}^{2}\right) $ is much less, than unity. It
means that space-time geometry is almost discrete. The quantity $1-\rho
\left( \sigma _{\mathrm{d}}\right) $, $\sigma _{\mathrm{d}}\in \left(
-\sigma _{0}-\lambda _{0}^{2},\sigma _{0}+\lambda _{0}^{2}\right) $ may be
interpreted as a degree of the discreteness of the space-time geometry. One
can see, that the discreteness of the space-time geometry and the degree of
the discreteness is determined by properties of the world function (but not
by properties of the manifold). The fact, that the space-time geometry,
given on a continuous manifold may be discrete, seems to be very unexpected.
This fact acknowledges the statement, that the world function and only world
function determines the space-time geometry.

It is reasonable to interpret the relative density $\rho \left( \sigma _{%
\mathrm{d}}\right) =d\sigma _{\mathrm{d}}/d\sigma _{\mathrm{E}}$ of points
of the distorted space-time with respect to the density of points of the
standard (Minkowskian) space-time as a measure of the space-time
granulation. Discreteness is a special case of graininess. Continuity is
another special case of graininess. The graininess of the space-time
describes also all intermediate cases, when the space-time is partly
continuous and partly discrete. Interrelation of the graininess with the
discreteness reminds interrelation of rational numbers with the natural ones.

Graininess of the space-time is a physical property of the space-time,
whereas the multivariance is a mathematical property of the space-time
geometry. The graininess of the space-time is connected with the
multivariance, and the world function formalism is a mathematical technique,
which can describe the graininess of the space-time.

One can easily imagine two limit cases of graininess: discreteness and
continuity. The relative density $\rho \left( \sigma _{\mathrm{d}}\right)
=d\sigma _{\mathrm{d}}/d\sigma _{\mathrm{E}}$ admits one to realize
intermediate cases of graininess. Conventional approach to geometry, based
on the concept of linear vector space, describes only continuous geometries.
Vector representation of geometry \cite{R2007}, based on the concept of the
linear vector space, cannot describe indefinite graininess of the
space-time. No finesses, based on the vector representation of geometry,
enables to describe effectively the space-time, whose graininess
distinguishes from continuity.

The discrete values of the elementary particles characteristics (mass,
charge, spin) are generated by some discrimination mechanism. The reason of
this discrimination is conditioned by the multivariance (more exactly by
zero-variance) of the space-time geometry \cite{R2007c}. From physical
viewpoint the reason of the discrete characteristics is the graininess of
the space-time.

\section{Concluding remarks}

Thus, the multivariance is a general property of the space-time geometry.
Class of uniform, isotropic flat space-time geometries is a continual set,
whose elements are labelled by a real function of one argument. Only one
geometry of this class may be considered as single-variant (geometry of
Minkowski). All other space-time geometries are multivariant. Considering
the Riemannian geometry as a more general space-time geometry, we restrict
our capacities. At the conventional approach, based on concepts of the
linear vector space, the natural multivariance of the Riemannian geometry is
suppressed by means of the fernparallelism interdict.

In the framework of the Riemannian geometry one cannot describe such
properties of the space-time as the limited divisibility and the graininess,
which are very important for geometrical description of elementary
particles. In general, ignoring multivariant geometries, we manifest, that
our knowledge of geometry are very restrictive. Our knowledge of geometry
does not admit one to construct effective description and effective dynamics
of elementary particles in microcosm, where the graininess of the space-time
is important.

The multivariance and the graininess are connected between themselves.
However, the graininess is rather physical concept, whereas the
multivariance is rather mathematical concept. The multivariance describes
the interrelation of two vectors, whereas the graininess describes
interrelation of the point density in the space-time with the standard point
density in the space-time of Minkowski. The graininess is more demonstrative
and more complicate, whereas the multivariance is less demonstrative and 
simpler. As a result the multivariance is considered as a basic concept of
the space-time, whereas the graininess is considered as a derivative concept.

In the Riemannian geometry the unlimited divisibility of the space-time
takes place. As a result, on one hand, the particle dynamics can be
described in terms of differential equations, whose application supposes the
unlimited divisibility of the space-time. On the other hand, the unlimited
divisibility generates such problems as confinement

One can hardly formulate mathematically the particle dynamics in the grainy
space-time, where the space-time divisibility is restricted, and one cannot
use differential equations. In the grainy space-time the particle dynamics
is determined by the space-time geometry in itself and by the structure of
the particle. Such a geometric dynamics is formulated in terms of the world
chain with finite links \cite{R2008}. The world chain is such a
generalization of the world line, when infinitesimal segments of the world
line are replaced by finite geometrical objects.

\end{document}